# Experimental Evaluation of Ventilation Systems in a Single-Family Dwelling

Juslin Koffi [1], Francis Allard [1] and Jean-Jacques Akoua [2]

[1] *Laboratoire d'Etude des Phénomènes de Transfert et de l'Instantanéité : Agro-industrie et Bâtiment (LEPTIAB), University of La Rochelle, Avenue Michel Crépeau 17042 La Rochelle Cedex 1, France*
[2] *Energy Division, Centre Scientifique et Technique du Bâtiment (CSTB), 84 Avenue Jean Jaurès Champs-sur-Marne 77447 Marne-la-Vallée Cedex 2, France*

## ABSTRACT

The French regulation on residential building ventilation relies on an overall and continuous air renewal. The fresh air should enter the building through the "habitable rooms" while the polluted air is extracted in the service rooms. In this way, internal air is drained from the lowest polluted rooms to the highest polluted ones. However, internal pressure equilibrium and air movements in buildings result from the combined effects ventilation system and parameters such as wind, temperature difference or doors opening. This paper aims to analyse the influence of these parameters on pollutant transfer within buildings. In so doing, experiments are carried out using tracer gas release for representing pollution sources in an experimental house. Mechanical exhaust, balanced and natural ventilation systems are thus tested. Results show the followings:
- *For all cases, internal doors' opening causes the most important pollutant spread.*
- *When doors are closed, the best performances are obtained with balanced ventilation.*
- *Independently to heating and doors opening,* pollutant transfer from lower levels to higher levels of the building is limited in the case of balanced ventilation.
- Pollutant spread is quite similar whether the whole building is heated or not.

## INTRODUCTION

Ventilation in buildings should fulfill three main objectives: i.e. to remove the indoor pollutants and prevent the outdoor pollutants from entering, protect building materials, and to control the indoor climate. For doing that first objective, the French regulation on residential building ventilation relies on a general and permanent air renewal of the building.

The fresh air should enter the building through the "habitable rooms" by natural air inlets or mechanical air supply. The polluted air is extracted in the service rooms (bathroom, shower, kitchen and toilets). In this way, internal air is drained from the lowest polluted rooms to the highest polluted ones. However, air and pollutant transfer in buildings is not only due to the ventilation system. Indeed, it also the consequence of the combination of parameters such as wind, temperature difference or doors opening, resulting in internal pressure equilibrium.

Air flow pattern in buildings is a parameter of interest. It is directly linked with indoor pollutant transfer as, most of the time, convective air flows taking place in the building remains the main factor of air change. To assess pollutant spread in building, numbers of works are done; however, a few experimental studies comparing different ventilation systems in a whole house are available. Afshari (Afshari, 2005) has experimentally compared three ventilation systems in four-room houses, but the results are focused on energy consumption. Koinakis (Koinakis, 2005) performed some experimental tests with natural ventilation systems in order to validate a numerical tool. Lin et al. (Lin, 2005) studied the effects of doors opening on the performance of displacement ventilation.

The present study aims to compare, by experimental means, the performance of three ventilation systems in terms of pollutant spread, i.e. to check out how a pollutant is distributed within a dwelling according. In this way, the influence of wind, temperature difference and internal doors on pollutant spread are evaluated for each ventilation system.

## MATERIALS AND METHODS

The experiments are carried out in the experimental house MARIA (i.e. *Mechanized house for Advanced Research on Indoor Air*) (Ribéron, 2001). This three-level full-scale house is located at the CSTB research centre near Paris. It consists of four bedrooms, a shower and a bathroom on first-floor. The living-room, the kitchen and the toilets are situated on ground-floor; there is a garage on the basement (Figure 1). The ground-floor is linked to the others by a stair; in addition, it is separated with the basement by a door. MARIA is built for studies in the fields of IAQ (Akoua, 2004), ventilation systems (Koffi, 2005), numerical tools validation (Koffi, 2009), etc.

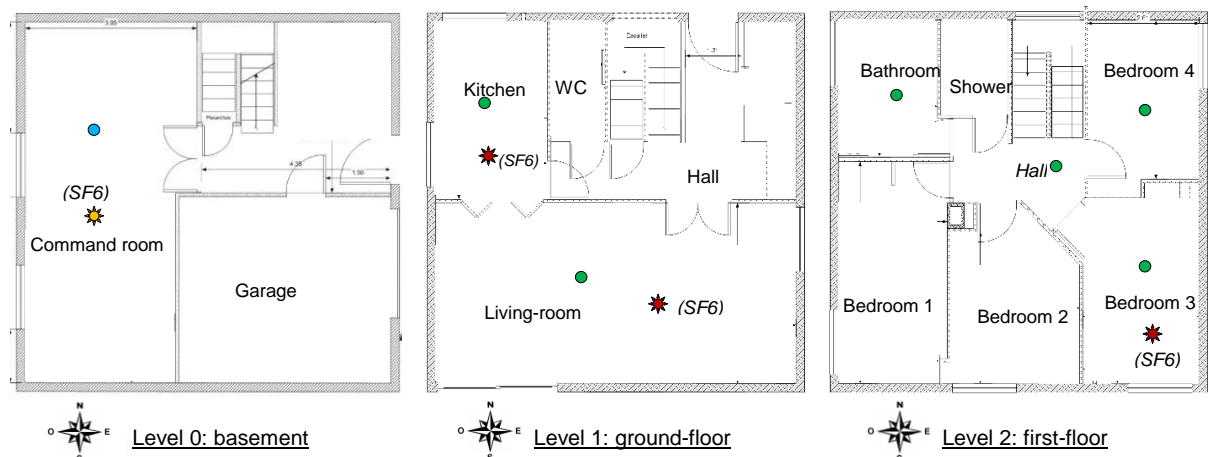

Figure 1: Tracer gas emission and concentration measuring positions in MARIA.

Tracer gas constant injection (Roulet, 1991) is carried out in order to simulate a pollutant emission (representing occupant presence, activity or material releases) respectively in the living-room, the kitchen, bedroom 3 and the basement: sulphur hexafluoride ($SF_6$) is released during 5 hours at 2 ml/s flow rate. Concentrations are measured for each case in the living, the kitchen, the bathroom, the hall, bedroom 3 and bedroom 4 or basement (Figure 1). Pollutant concentrations in the bedrooms aim to assess airflow transfer within the dwelling. At the end of the injection, decay method is carried out to estimate air change rate in the source room. These measurements are done with a six-point sampler for pumping air samples from measurement points, and a multi-gas monitor that evaluates the concentrations.

This method is used to compare the performance of three ventilation systems installed in MARIA and commonly used in French residential buildings: mechanical exhaust ventilation, balanced ventilation (mechanical exhaust and supply with heat recovery) and natural ventilation. The latter system is made of individual vertical ducts in service rooms. Two pressure-controlled air inlets (30 m$^3$/h at 20 Pa) are installed in the living-room and one in each bedroom for both natural and exhaust ventilation systems. The extracted airflow rates in each exhaust room are quite similar for mechanical ventilation systems. In case of balanced ventilation only, the total supply airflow rate is slightly greater than the exhaust airflow rate.

Tests are first performed for each ventilation system with doors closed and no heating; after that, the heating system is switched on. The same procedure is performed when internal doors are opened. The following results focus on doors and wind effects.

## RESULTS

### Comparison of Ventilation Systems

Figure 2 through Figure 4 present the results of emissions in the living-room for the studied ventilation systems. Each figure deals with a ventilation system. One can see the evolution of tracer gas concentrations in the living-room and the spread in the kitchen, the hall, bedrooms 3 and 4, and the bathroom.

Concentration in the injection room is found higher in the case of natural ventilation and lower for balanced ventilation. In spite of that, transfer to the kitchen quite is similar for both mechanical ventilation systems. In addition, concentrations in the hall and the bathroom are also similar for these systems. These results are due to comparable exhaust airflows in service rooms.

However, pollutant transfer to bedrooms shows differences. Almost no pollution is measured in these rooms for balanced ventilation (Figure 4). In fact, air supply in bedrooms combined with extraction in service rooms drain correctly internal airflows from main rooms to service rooms. In case of exhaust ventilation, bedroom 4 shows a higher pollution than bedroom 3 (Figure 3). There may be an inversion of air flow pattern between bedroom 4 and the hall, certainly resulting of wind and thermal buoyancy effects.

In the case of natural ventilation (Figure 2), internal air flows are only due to wind and temperature difference between the dwelling and outdoor. There is no mechanical air diffusion system. Because the experiments are carried out at the beginning of the hot season, temperature differences are quite low, so there is an accumulation of the tracer gas within the dwelling. Thus, the extraction is more submitted to wind effects on the building. For instance, concentrations in the bedrooms exceed 100 mg/m$^3$ while they are lower than 50 mg/m$^3$ for exhaust ventilation.

For an injection in the kitchen, the tracer gas is completely extracted by mechanical systems, no pollution is found elsewhere. However, in case of natural ventilation, the extraction in the kitchen is not enough sufficient. Then, the measures show a great spread of the tracer gas in the living-room and at the first level.

When the pollution takes place in bedroom 3 (1st level), only a few proportion of the pollutant can reach the ground-level. This result is due to the internal thermal buoyancy that causes a rise of air from lower levels to higher ones in building. However, bedroom 4 nearby is sometimes polluted, especially for natural and exhaust ventilation system.

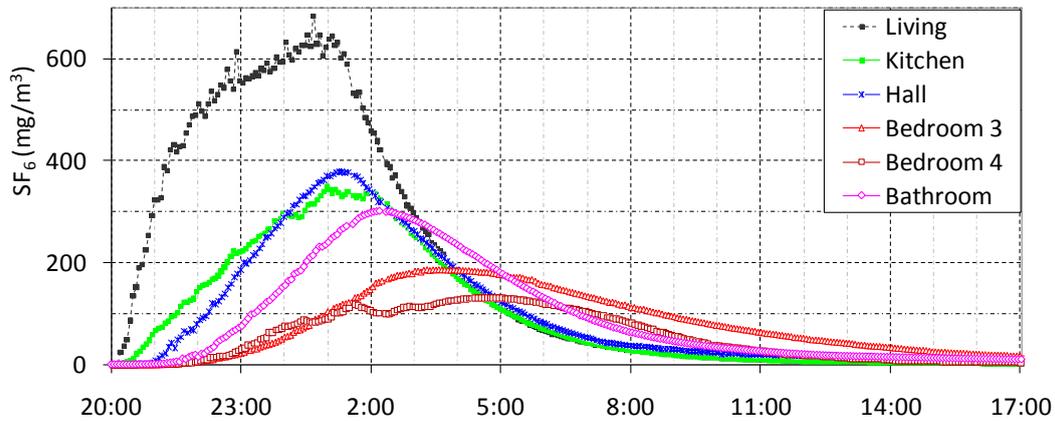

Figure 2: Concentrations for injection in the living-room with natural ventilation.

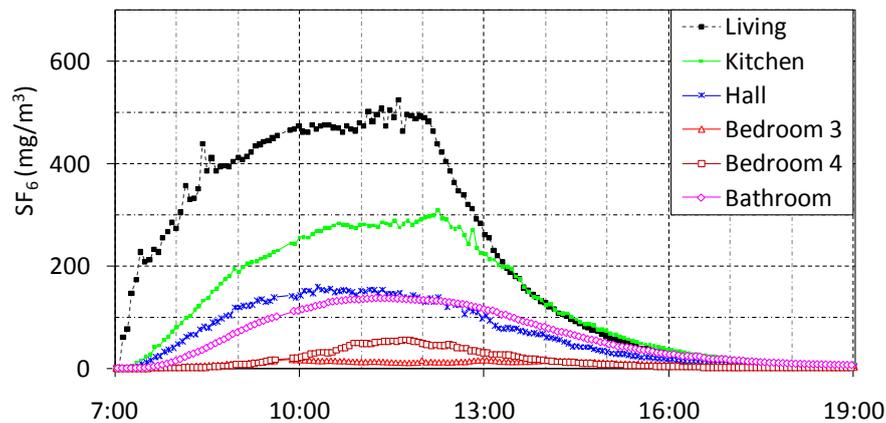

Figure 3: Exhaust ventilation - concentrations for injection in the living-room.

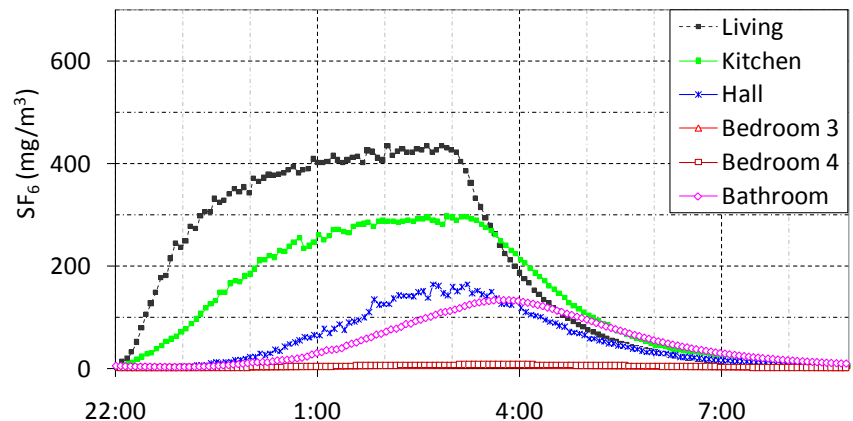

Figure 4: Balanced ventilation - concentrations for an injection in the living-room.

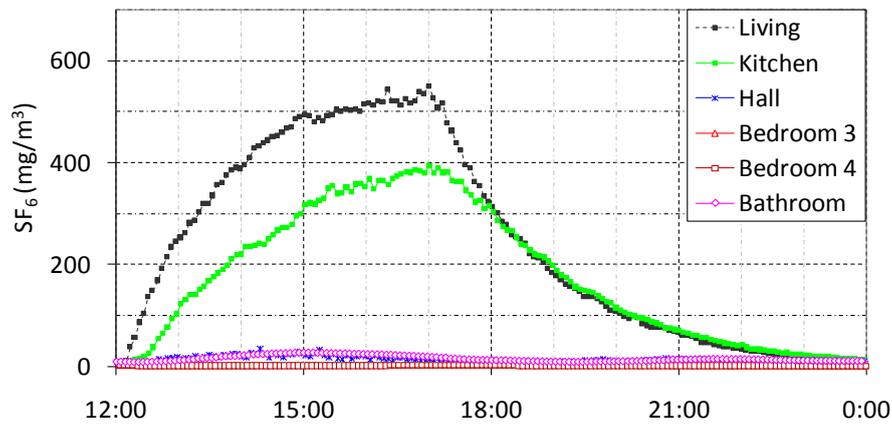

Figure 5: Balanced ventilation - wind effects on tracer gas transfer.

These differences of performances mainly result from air renewal differences due to the systems themselves but also to wind effect.

**Wind Effects**

Wind effects on internal airflows are illustrated by Figure 5 for balanced ventilation where the concentrations the hall and the bathroom are lower than those presented by Figure 4. In the case of Figure 5, the façade of the hall is situated upwind while the living-room is downwind. The infiltration through the hall façade cut-circuits internal air flows towards the first level of the building and creates a depression in the living-room. This results in very low gas concentration at the first level and an accumulation of the pollutant in the living-room. Nevertheless, almost the pollutant is extracted in the kitchen. For natural ventilation and exhaust ventilation, infiltration due to wind leads to lower or higher pollution of bedrooms compared to the results presented respectively by Figure 2 and Figure 3, depending on the direction of wind with regard to these rooms.

**Effects of Doors Opening**

Figure 6 shows internal doors effects on concentrations for balanced ventilation.

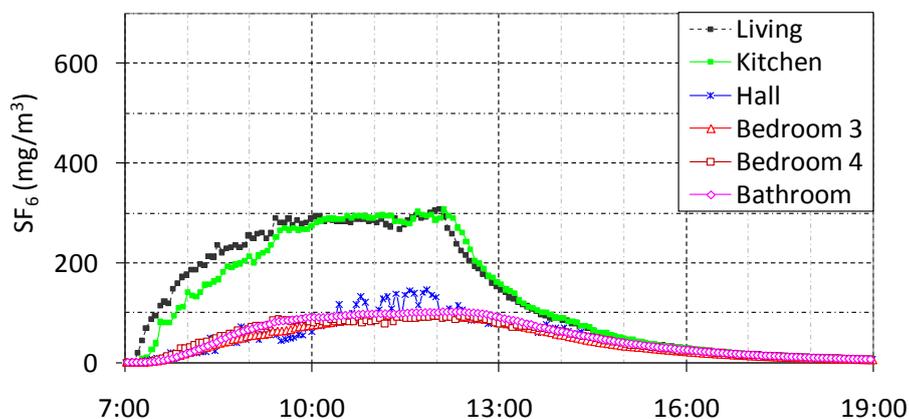

Figure 6: Balanced ventilation - internal doors opening effects on tracer gas transfer.

Independently to the ventilation system and the location of the pollutant source, doors opening results in a high pollution of the whole building.

Measured concentrations are homogenous at each level of the house. Airflow through doorways occurs in two directions, so that it promotes a "good" mixing of air in the kitchen and the living-room as they are separated by an open door. The same phenomena occur upstairs. But it appears that the pollution is higher at ground-level and lower at first level with balanced ventilation compared to exhaust system. This is again due to the difference between these systems, i.e. controlled air supply in the main rooms.

**Case of the Basement**

When the pollution takes place in the basement, the results depend on the ventilation system. On the one hand, no pollution is found in the habitable levels of the dwelling with balanced ventilation. In fact, the equilibrium between air extraction and air supply results in a zero-pressure zone in the hall, so that airflow from the basement to the hall is very low. On the other hand, the depression created by exhaust ventilation leads to the pollution of the higher levels of the dwelling, as the airflow from the basement to the hall becomes greater. Even all tested main rooms are polluted.

**CONCLUSION**

The experiments carried out in MARIA show many results with regards to wind, internal doors and ventilation systems. Airflows are well drained towards service room when the house is equipped with balanced ventilation due to air supply in bedrooms and living-room. The performance of natural ventilation is limited by lower temperature differences combined with wind infiltration: all rooms are polluted by tracer gas. For exhaust ventilation, internal airflows are subjected to wind effects resulting in relative high pollution of rooms situated downwind.